\documentclass[aps,preprintnumbers,prl,twocolumn,superscriptaddress]{revtex4}
\usepackage{epsfig,latexsym,cancel,amssymb,amsmath}
\usepackage{graphicx}
\usepackage{epstopdf}
\usepackage{color}
\usepackage{slashed}
\usepackage{ulem}

\def\beq{\begin{equation}}
\def\eeq{\end{equation}}
\def\bea{\begin{eqnarray}}
\def\eea{\end{eqnarray}}

\begin{document}

\title{Disentangling Strong Dynamics through Quantum Interferometry}

\author{Hitoshi Murayama}
\affiliation{Department of Physics, University of California, Berkeley  85721, USA}
\affiliation{Kavli Institute for the Physics and Mathematics of the Universe (WPI), Todai Institutes for Advanced Study, University of Tokyo, Kashiwa 277-8583, Japan}
\affiliation{Theoretical Physics Group, Lawrence Berkeley National Laboratory,
  Berkeley, CA 94720, USA}
\author{Vikram Rentala\footnote{Current address: Department of Physics, Indian Institute of Technology - Bombay, Mumbai 400076, India}}
\affiliation{Department of Physics \& Astronomy, Michigan State University, E. Lansing, MI 48824, USA}
\author{Jing Shu}
\affiliation{State Key Laboratory of Theoretical Physics and Kavli Institute for Theoretical Physics China (KITPC), Institute of Theoretical Physics, Chinese Academy of Sciences, Beijing 100190, P. R. China}

\begin{abstract}
We present a new probe of strongly coupled electroweak symmetry breaking at the 14~TeV LHC by measuring a phase shift in the event distribution of the decay azimuthal angles in massive gauge boson scattering. One generically expects a large phase shift in the longitudinal gauge boson scattering amplitude due to the presence of broad resonances. This phase shift is observable as an interference effect between the strongly interacting longitudinal modes and the transverse modes of the gauge bosons. We find that even very broad resonances of masses up to 900~GeV can be probed at 3$\sigma$ significance with a 3000~fb$^{-1}$ run of the LHC by using this technique. We also present the estimated reach for a future 50 TeV proton-proton collider.
\end{abstract}

\maketitle

%\section{Introduction}

\noindent{\bfseries Introduction.} One of the most important goals of the Large Hadron Collider (LHC) is to find the nature of the mechanism of electroweak symmetry breaking (EWSB). The discovery of the 125 GeV Higgs-like object \cite{Aad:2012tfa, Chatrchyan:2012ufa} is a major milestone in this direction. However, in order to truly understand EWSB, we would like to learn the origin of the longitudinal components of massive electroweak gauge bosons ($V$). Broadly speaking, models of EWSB fall into two categories, those where the longitudinal components of the $V$s are a) weakly interacting or b) strongly interacting. The most popular examples in the first category are the Standard Model (SM), with one or more elementary Higgs multiplets \cite{Branco:2011iw}, and its supersymmetric extensions \cite{Martin:1997ns}. In the second category, one or more strongly interacting sectors appear at the TeV scale and are responsible for EWSB. The Higgs-like boson and the longitudinal $V$s could arise as pseudo-Nambu-Goldstone bosons (PNGBs) in this scenario \footnote{See Ref. \cite{Contino:2010rs} and references therein.}. A definite distinction between these two cases at the LHC would be very important and serve as one of the first crucial steps towards a full understanding of EWSB mechanism.

One universal consequence of a strongly coupled EWSB sector is an enhancement of the longitudinal gauge boson scattering amplitude at high energies over the standard model expectation \footnote{There could be other non-universal consequences, such as resonances, enhanced di-higgs production rate \cite{Contino:2010mh}, etc.}. In order to measure such an effect, one would look for an enhanced rate for total gauge boson scattering at high invariant mass for the gauge boson pair, and then use the \textsl{polar} angle distribution of the gauge boson decay products to measure their longitudinal polarization fraction \footnote{See Ref. \cite{Han:2009em} and references therein.}.

In addition to an enhancement of the \textsl{magnitude} of the amplitude, one also expects a large \textsl{phase shift} in the amplitude, as in the case of scattering through a resonance. In this paper, we would like to seek strategies to experimentally probe this phase shift induced as a consequence of strong dynamics. We will use the \textsl{azimuthal} angle correlations of the $V$s' decay products and show that the strong phase shift shows up as a modification to the interference effect between the longitudinal and transverse $V$ polarizations in the azimuthal angle distributions. Thus, the phase shift is turned into an observable that can be used as a complementary probe (in addition to the longitudinal V scattering rate) of EWSB from strong dynamics.

\begin{figure}
\setlength{\belowcaptionskip}{0pt}
\begin{center}
  \includegraphics[width=8cm]{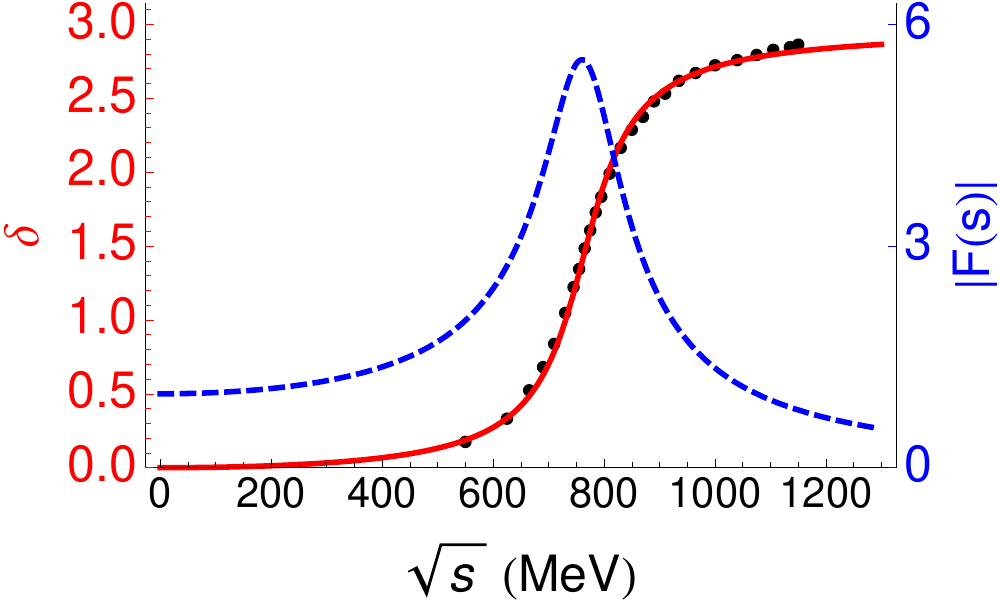} \\
\end{center}
  \caption{The phase shift $\delta$ in $ e^+ e^- \rightarrow \pi \pi$ scattering versus energy $\sqrt{s}$. The experimental data (black points) are from Ref. \cite{Protopopescu:1973sh}. The red solid line is the ansatz in Eq. (\ref{eq:phase}) we have used to fit the phase shift using the parameters $m = 760$ MeV and $\Gamma= 140$ MeV. Using dispersion relations we can relate the phase shift to the magnitude of the form factor $|F(s)|$. For $P(s) =1$, the magnitude is shown by the blue dashed line.}
    \label{fig:ffphase}
\end{figure}

\noindent{\bfseries PNGB scattering and parameterization of the phase shift.} Pion scattering provides a realistic example of strongly coupled PNGB scattering. By looking at the experimental $\pi \pi $ scattering data \cite{Protopopescu:1973sh}, we can see that a large phase shift exists in the form-factor of both $\pi \pi \rightarrow \pi \pi$ and $e^+ e^- \rightarrow \pi \pi$. The phase shift $\delta$ in $ e^+ e^- \rightarrow \pi \pi$ scattering versus energy $\sqrt{s}$ is shown by the black data points in Fig.~\ref{fig:ffphase}. We can see that the amplitude undergoes a large phase shift when $\sqrt{s}$ is near the mass of the $\rho$ meson (760~MeV).

The low energy effective theory of pion scattering is described by a chiral Lagrangian. In this description, the scattering amplitude in the isospin $(I)$ and partial wave $(J)$ channel of $\pi \pi \rightarrow \pi \pi$ is given by,
\begin{equation}
\mathcal M_{IJ} \propto \sin(\delta_{IJ}) e^{i \delta_{IJ}} \simeq c_{IJ} \frac{s}{f_{\pi}^2} e^{ i s/f_{\pi}^2},
\end{equation}
where $f_\pi = 84 \textrm{ MeV}$ is a low-energy constant of chiral perturbation theory (which differs from the physical pion decay constant, $f^\textrm{phys}_\pi = 92 \textrm{ MeV}$ that arises away from the chiral limit), and the last equality follows from the low energy theorem prediction of the behavior of the scattering amplitude with $s$, which also determines the constants $c_{IJ}$. Naive extrapolation of this form of the amplitude violates unitarity for energies $\sqrt{s}$ comparable to the cutoff $4\pi f_{\pi}$ of the effective theory.
In particular the low energy theory fails to predict a resonant enhancement of the magnitude of the amplitude, as well as the large phase shift expected from the exchange of ($J=1$) vector-meson resonances.

There are several ways to incorporate the effect of the vector resonances into the scattering amplitude calculations of the low energy theory. One approach is to add in a broad vector resonance by hand to the theory. However, we will choose to adopt a different approach that manifestly maintains the unitarity of the amplitude at very high energies and emphasizes the central role of the phase shift.

We will choose to multiply the tree level amplitude (in the $J=1$ channel) by a complex form factor $F(s)$. The entire form factor can then be extracted from its phase by using analyticity arguments to define an Omn\`{e}s function \cite{Omnes:1958hv} and assuming no inelastic channels. For a given form factor, $F(s)$, applying the subtracted dispersion relation to $\log(F(s)) / s$, we have
\bea
\label{eq:Omnes}
F(s) = P(s) \exp [ \frac{1}{\pi} \int_0^{\infty} d s' \delta(s')  \{ \frac{1}{s' - s - i \epsilon} - \frac{1}{s'} \} ] \ .
\eea
When the phases $\delta(s')$ goes beyond $2 \pi$ (for instance, multiple resonances with additional branches), the additional $2 \pi n$ phase factors can be recast into $P(s)$ as a polynomial factor with $P(0)=1$.

For simplicity, we use an ansatz for the phase,
\bea
\label{eq:phase}
\delta(s) = \left\{
\begin{array}{lr}
\textrm{ArcTan}[s \Gamma / m(m^2 + \Gamma^2 -s)], & s<m^2 \\
\textrm{ArcTan}[s \Gamma / m(m^2 + \Gamma^2 -s)] + \pi, & s \geqslant m^2,
\end{array} \right.
\eea
which approaches a constant at high energy according to unitarity.
We can see that this ansatz can fit the phase of the $\pi \pi $ scattering data very well, as shown in Fig. \ref{fig:ffphase}.  Nevertheless, our parametrization is general and the does not rely on the specific form of strong dynamics. From Eq. (\ref{eq:Omnes}), we can construct the full form factor from its phase $\delta(s)$,
\bea
\label{eq:formfactor}
F\left( s \right)= P(s) \frac {-{m}^{{2}}+i m\Gamma }{s-{m}^{{2}}+i m\Gamma }. \
\eea
We can see the behavior of $|F(s)|$ in Fig. \ref{fig:ffphase}. At small $s$ we are far from the resonance and the form factor is unity. However as we approach the resonance the magnitude of the form factor grows large. For large values of $\sqrt{s}$ beyond the resonance, the form factor falls off rapidly with energy.

\noindent{\bfseries Longitudinal weak bosons scattering.}
In a strongly interacting EWSB theory the longitudinal components of the weak bosons can be approximately regarded as PNGBs, and their interactions can be described at low energies by a chiral Lagrangian. Thus, naive extrapolation of the scattering amplitude would lead to a similar problem of unitarity violation. We can expect a similar resolution to this problem as in the case of pions, where new resonances unitarize the scattering amplitude.

However, there is one key difference between weak boson scattering and pion scattering. Namely, we have already discovered a 125 GeV (Higgs-like) scalar object that couples to weak bosons. The exchange of this scalar object will partially restore unitarity in the longitudinal $V$ scattering amplitude. If the couplings of this scalar object to $V$s is exactly the Standard Model value, corresponding to a non-composite Higgs boson, then we would have complete unitarity restoration just from exchange of this scalar object.

For composite Higgs models, we can use the generalized Adler-Weinberg sum rule (in the limit of vanishing gauge couplings) \cite{Falkowski:2012vh} to relate the Higgs coupling to $V$s to an integral sum of longitudinal gauge boson scattering cross-sections in various isospin channels,
\bea
1-a^2 = \int_0^\infty \frac{v^2 ds}{6 \pi s} (2 \sigma^{\textrm{tot}}_{I=0}(s) + 3 \sigma^{\textrm{tot}}_{I=1}(s) - 5 \sigma^{\textrm{tot}}_{I=2} (s) ).
\eea
Here, $a$ parameterizes the ratio of the Higgs boson coupling to $V$s over its SM value. From the latest fits \cite{Giardino:2013bma} we can see that $0.8<a<1.2$ at the 2$\sigma$ level.

For $a$ not equal to 1, we need additional contributions from strong dynamics to restore unitarity in longitudinal gauge boson scattering. For simplicity, we will also assume vector meson dominance for the remaining partial unitarity restoration, as in the case of pion scattering, restricting the new physics contribution to the $J=1$ channel.

We can now parameterize strong dynamics in longitudinal $V$ scattering by introducing a form factor that multiplies the amplitude just as we did in the case of $\pi\pi$ scattering. We will assume the same ansatz for the phase, since it maintains unitarity manifestly. The role of the composite Higgs boson in unitarization can be accounted for without explicitly including the Higgs exchange diagrams. Instead, we note that $J=1$, longitudinal $V$ cross-section must be rescaled by a factor of $1-a^2$ (compared to the case where no Higgs-like boson is present). Thus, we can simply rescale our form factor from Eq. (\ref{eq:formfactor}) by $\sqrt{1-a^2}$ to account for the presence of the 125 GeV boson.

\begin{figure}
\setlength{\belowcaptionskip}{5pt}
\begin{center}
  \includegraphics[width=8cm]{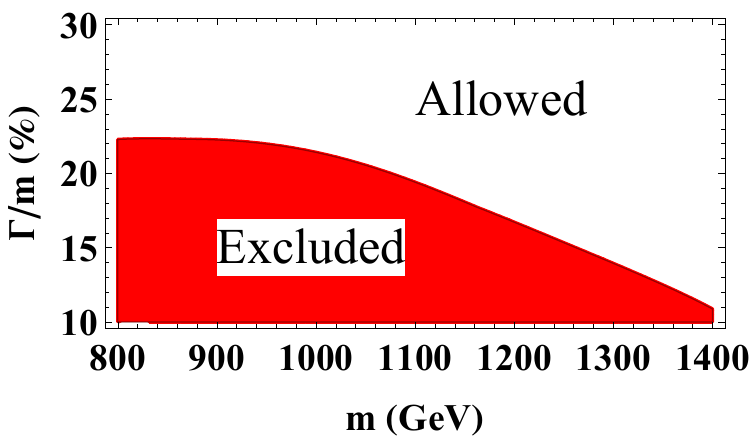}
% \newpage
\end{center}
  \caption{The currently excluded region of form-factor parameters (shaded region) using constraints from CMS searches for $W^\prime \rightarrow W Z$ resonances \cite{WZbound}. (The ATLAS bound is slightly weaker and is not shown.) }
  \label{fig:WZbound}
\end{figure}

Assuming $a = 0.8$, we show the currently excluded region for different form-factor parameters $m$ and $\Gamma$ in Fig.~\ref{fig:WZbound} using the latest LHC searches for $W^\prime \rightarrow W Z$ resonances \cite{WZbound}. This search strategy only probes the enhancement to the total rate for $WZ$ production, or equivalently it is a probe of $|F(s)|$, but it does not probe the phase shift directly. In addition, the search uses a narrow $WZ$ invariant mass window to suppress backgrounds and is therefore insensitive to very broad resonances $\Gamma/m \gtrsim 20\%$, that are characteristic of strong dynamics.

\noindent{\bfseries Observation of Longitudinal V scattering.}
There are two methods to observe $VV$ scattering at the LHC. One is the widely used weak boson fusion process $pp \rightarrow VVjj$, where the two forward jets can be used to suppress the large SM backgrounds. The other method is the rescattering process $pp \rightarrow VV$, which is not considered to be a promising channel for most studies due to the large SM backgrounds. However, this channel is not suppressed by the small effective-$V$ luminosity and it also has better access to higher energies of the $VV$ system. Since we are trying to observe the azimuthal angle correlation which arises from a quantum interference term, we do not have to suppress the SM backgrounds and we will consider the rescattering process in this paper.

%\section{Parametrization}
%\smallskip

\begin{figure}
\setlength{\belowcaptionskip}{0pt}
\begin{center}
  \includegraphics[width=8cm]{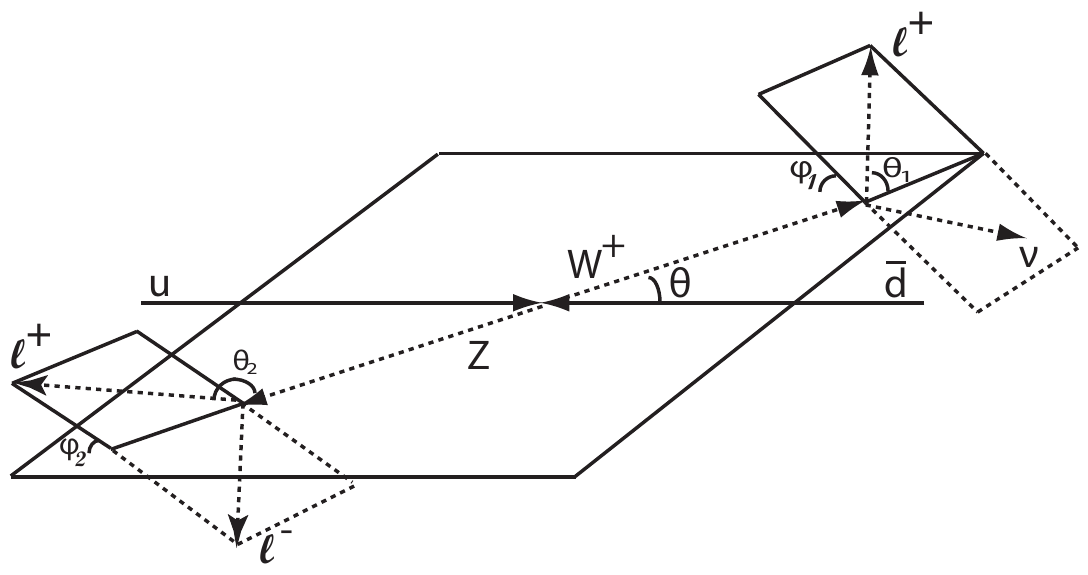} \\
% \newpage
\end{center}
  \caption{The kinematics of $W^+ Z$ production and decays.}
  \label{fig:angles}
\end{figure}

\smallskip
\noindent{\bfseries Angular correlation.} Let us consider $W^+ Z$ production from a $u \overline{d}$ initial state. We will consider a modification of the SM amplitude of longitudinal $W^+Z$ production in the $J=1$ channel by a complex form-factor $\sqrt{1-a^2} F(s)$, where $F(s)$ is given by the form in Eq. (\ref{eq:formfactor}). We will study this process at high energies  where both the $W^+$ and the $Z$ decay leptonically. The kinematic dependence of the production and decay amplitudes are as follows: $\mathcal{M}_1: u(k_1, -) ~ \bar{d} (k_2, +)  \rightarrow W^+ (q_1, \lambda_1) ~ Z (q_2, \lambda_2)$, $\mathcal{M}_2: W^+ (q_1, \lambda_1) \rightarrow  \nu (p_1, -) ~ l_1^+ (p_2, +)$ and $\mathcal{M}_3: Z (q_2, \lambda_2) \rightarrow l_2^- (p_3, h) ~ l_3^+(p_4, - h)$.  The parameters in the parentheses are the particle momentum and helicity respectively. The phase space of this process has five independent angles which are defined in the center-of-momentum frame as follows: the production angle ($\Theta$), two polar decay angles ($\theta_1$, $\theta_2$) and two azimuthal decay angles ($\phi_1$, $\phi_2$) which can be though of as the rotations of the $W^+/Z$ decay planes ($\hat{n}_W$,$\hat{n}_Z$) about the $W^+/Z$ momentum axis, and are measured relative to the production plane $\hat{n}$. The three planes are defined as $\hat{n} \sim k_1 \times q_1$, $\hat{n}_W \sim q_1 \times p_2$ and $\hat{n}_Z \sim p_4 \times q_2$. All these kinematic variables are presented in Fig. \ref{fig:angles}.

The phase shift from the strong dynamics only affects the longitudinal-longitudinal combination of $W^+Z$ modes and shifts the corresponding amplitude by an energy dependent phase, $\delta$. This phase shift will enter into the azimuthal angle correlation of the $W^+/Z$ decay as an interference effect between the various polarizations of the vector bosons. To see this, recall that $W^+/Z$ decay produces an azimuthal angular dependence $\exp(i s_z \phi)$ in the amplitude, where $s_z$ and $\phi$ are respectively, the spin projection and the azimuthal angle rotation about the $W^+/Z$ direction. For a given interference term between a general helicity combination ($\lambda_1, \lambda_2$) and the (0,0) combination, we find that the relevant terms in the differential cross-section are of the form $ \left|a_1 e^{i (\lambda_1 \phi_1 -\lambda_2 \phi_2)} + a_2 e^{i \delta} \right|^2 \sim  \cos(\lambda_1 \phi_1 -\lambda_2 \phi_2 +\delta) \supset \sin(\lambda_1 \phi_1 - \lambda_2 \phi_2) \sin \delta$ \cite{Keung:2008ve,Cao:2009ah}. Thus, the $\sin(\lambda_1 \phi_1 - \lambda_2 \phi_2)$ modes in the azimuthal angle correlation strongly suggest the existence of strong dynamics.
%The $\cos \phi$ and $\cos 2\phi$ modes can be used to study the W/Z spin \cite{arXiv:0804.0476}.

%${\cal M}_{\lambda_1 \lambda_2}$ and $ {\cal M}_{0,0}$

%The full production amplitude for different $W^+ Z$ helicity combinations ($\lambda_1, \lambda_2$) are presented as a function of $\cos\Theta$ in Fig. \ref{fig:production}.
The production amplitudes can be separated by the different $W^+ Z$ helicity combinations ($\lambda_1, \lambda_2$). Among the nine different helicity combinations, there are three leading contributions from $(-, +)$, $(0,0)$ and $(+,-)$. The four other significant ones are $(-, 0) \approx (0,+)$ and $(+, 0) \approx (0, -)$, all of which have a relative suppression $\sim m_W/\sqrt{s}$ compared to the leading modes. The $(+, +)$ and $(-,-)$ combinations are too small to affect the kinematic distributions. We note that (a): $(-,+)$/$(+,-)$ dominate as $\Theta \rightarrow 0$ or $\pi$ because of the t-channel production. (b): Numerically the difference between $(-, 0)$ and $(0,+)$ or $(+, 0)$ and $(0, -)$ is negligible. We parameterize the production amplitudes as $\mathcal{M}_1 (-, +) = A$, $\mathcal{M}_1 (0, 0) = B e^{i \delta}$, $\mathcal{M}_1 (+, -) = C$, $\mathcal{M}_1 (-, 0) = D$, $\mathcal{M}_1 (+,0) = E$, $\mathcal{M}_1 (0,+) = F$ and $\mathcal{M}_1 (0,-) = G$. Here, all amplitudes depend on the center-of-mass energy and on $\Theta$. Similar behavior has been pointed out in $e^+ e^- \rightarrow W^+ W^-$ \cite{Hagiwara:1986vm}.

The full differential cross-section can be obtained from $\sum_{h}
 \left| \sum_{\lambda_1, \lambda_2}
\mathcal{M}_1^{\lambda_1 \lambda_2} (\Theta) \,
\mathcal{M}_2^{\lambda_1} (\theta_1, \phi_1)\,
\mathcal{M}_3^{\lambda_2} (h, \theta_2, \phi_2)\right|^2$, where the $W^+, Z$ decay amplitudes are $\mathcal{M}_2^{\lambda_1} (\theta_1, \phi_1)$ $= g_W |q_1| d_{\lambda_1}(\theta_1) e^{i \lambda_1 \phi_1}$, $\mathcal{M}_3^{\lambda_2} (h, \theta_1, \phi_1) = g_Z^h |q_2| d_{\lambda_2}(\theta_2 + (h-1) \pi /2) e^{- i \lambda_2 \phi_2}$.
%\blue{We use the approximation $g_Z^+ \approx - g_Z^-$ for the $Z$ charged leptonic decays.}
Here $g_Z^- \approx - g_Z^+$ are the couplings of the $Z$ to different lepton helicities.
The polar angle dependent function $d_\pm (\theta) = \sqrt{\frac{1}{2}} ( 1 \pm \cos \theta)$, $d_0 (\theta) = \sin \theta$. If we integrate over both polar angles $\theta_{1, 2}$, there is an approximate cancellation in the $\sin (\phi_1+\phi_2)$ and $\sin(\phi_2)$ correlation between $\cos \theta_2>0$ and $\cos \theta_2<0$ due to $g_Z^+ \approx - g_Z^-$. Therefore, we only integrate the differential cross-section over either $\cos \theta_2>0$ or $<0$ to obtain:

%\begin{widetext}
%\bea \frac{(2\pi)^2 d \sigma _{\pm}}
%{ d\cos\Theta  d\phi_1 d \phi_2} \over \frac{d\sigma}{d\cos\Theta}
%&=&{1\over 2}\left [
%\left (1 \pm \frac{3 \epsilon}{4} \frac{(C^2-A^2+G^2-F^2)}{H^2} \right)\right. %\nonumber \\ &&
% + \frac{B \sin \delta}{H^2} \left ( {3 \pi \over 64} \left ( \pm 4 (A+C) - 3 \pi \epsilon (A-C) \right )   \right.
%\sin(\phi_1 +\phi_2)
% \nonumber\\ &&
% +{3 \sqrt{2} \pi (E-D)  \over 8}  \sin \phi_1
%\nonumber\\ &&
%\left. \left.
%+ \frac{\pm 4(F+G) - 3 \pi \epsilon (F-G) }{4 \sqrt{2}} \sin \phi_2
%+ \cdots
% \right ) \right] \ ,
% \label{eq:diffdist}
%  \eea
%\end{widetext}
\begin{widetext}
\bea \frac{(2\pi)^2 d \sigma _{\pm}}
{ d\cos\Theta  d\phi_1 d \phi_2}
&=&{1\over 2}\left [
\left (H^2 \pm \frac{3 \epsilon}{4} (C^2-A^2+G^2-F^2) \right)\right. %\nonumber \\ &&
 + B \sin \delta \left ( {3 \pi \over 64} \left ( \pm 4 (A+C) - 3 \pi \epsilon (A-C) \right )   \right.
\sin(\phi_1 +\phi_2)
 \nonumber\\ &&
 +{3 \sqrt{2} \pi (E-D)  \over 8}  \sin \phi_1
%\nonumber\\ &&
\left. \left.
+ \frac{\pm 4(F+G) - 3 \pi \epsilon (F-G) }{4 \sqrt{2}} \sin \phi_2 \right )
+ \cdots
  \right] \ ,
 \label{eq:diffdist}
  \eea
\end{widetext}
where $\epsilon =((g_Z^-)^2 - (g_Z^+)^2)/((g_Z^-)^2 + (g_Z^+)^2) \approx 0.22$ and the ellipsis refer to the interference terms with $\cos \phi$-type dependence. The $\sigma_\pm$ stands for $\sigma (\cos\theta_2  {\ }^>_< \ 0)$ and $H$ is overall background $H^2 =  (A^2 +B^2 +C^2+D^2 +E^2+F^2 +G^2)$.

\begin{figure}
\setlength{\belowcaptionskip}{5pt}
\begin{center}
  \includegraphics[width=8cm]{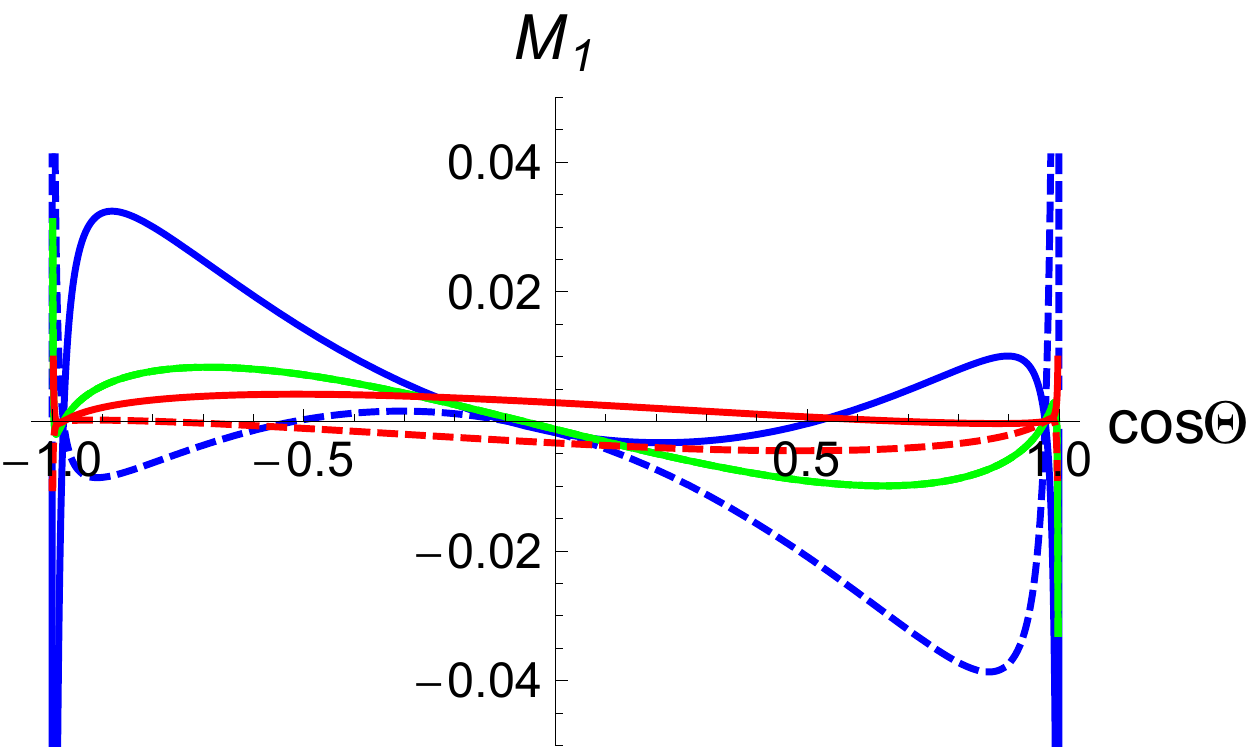}
% \newpage
\end{center}
  \caption{The coefficients of the production amplitudes in front of the azimuthal $\sin \phi$ correlations as a function of scattering angle $\cos \Theta$ at $\sqrt{s} = 1$ TeV from Eq. (\ref{eq:diffdist}). The solid blue, green, red and dashed blue, red lines stand for the coefficients in front of $\sin(\phi_1 + \phi_2)$ ($\cos \theta_2 >0$), $\sin\phi_1$, $\sin\phi_2$ ($\cos \theta_2 >0$), $\sin(\phi_1 + \phi_2)$ ($\cos \theta_2 <0$), $\sin\phi_2$ ($\cos \theta_2 <0$) respectively. }
  \label{fig:coefficients}
\end{figure}

In Fig.~\ref{fig:coefficients} we plot the relative coefficients of the different $\sin \phi$ correlations as a function of $\cos \Theta$ for $\sqrt{s} = 1$~TeV using the expressions in Eq.~(\ref{eq:diffdist}). Measuring a non-zero coefficient for any of the $\sin \phi$ modes is a positive indicator of strong dynamics. Integrating over the entire $\cos \Theta$ range would lead to cancellations that would dilute the significance of the probe. Thus, an optimal strategy is to use a maximum likelihood analysis on the measured $d\sigma_\pm/d\cos\Theta d\phi_1 d\phi_2$ distribution in the data to look for all non-zero $\sin(\lambda_1 \phi_1 - \lambda_2 \phi_2)$ coefficients.

%\begin{itemize}
%\item $\Delta r > 0.4$ separation between leptons.
%\item $p_T > 20 $ GeV and $ |\eta| < 2.8$ cuts on the leptons.
%\item $Z$-reconstruction cut: We require that two opposite sign leptons reconstruct to give the Z mass. %\blue{What is the $Z$ mass window??}
%\item Missing $E_T$ cut $> 20$ GeV. %( This was not impose in the code... )
%\item Invariant mass cut of the $W/Z$ system between $m-\Gamma$ and $m + \Gamma$. % to maximize the phase shift while still allowing for a large number of events
%\item  $0.1<\cos \Theta<0.9$ on the W momentum in the CM frame to maximize the interference. %(maybe show this plot).
%The upper cut of 0.9 kills the large $(+,-)$ and $(-,+)$ helicity combinations.
%\end{itemize}

%\begin{itemize}
%\item The $u$-quark direction is unknown.
%\item There is a two fold ambiguity in the neutrino momentum along the beam axis.
%$W^+$ has transverse momentum purely in the $+x$ direction.
%\end{itemize}

%\section{Procedure}
\smallskip
\noindent{\bfseries Procedure.}
At the LHC, there are various kinematic ambiguities that must be incorporated into the analysis. We will choose to study only the $\sin \phi_1$ mode, which will yield a much more transparent analysis that is robust to the kinematic ambiguities.

We simulate the process $p p \rightarrow W^+ Z \rightarrow l^+ \nu l^+ l^-$ with the form factor $\sqrt{1-a^2} F(s)$ from Eq. (\ref{eq:formfactor}) multiplying the $J=1$ channel of the $(0,0)$ helicity amplitude. We have taken $a=0.8$ as a benchmark, which assumes that such a deviation in Higgs couplings would continue to be allowed with future data from the LHC. We then scan over different form factor parameters $m$ and $\Gamma$, which will give rise to different phase shifts.
 %We use the HELAS \cite{Murayama:1992gi} program to simulate events at the parton level which is sufficient for the fully leptonic final state.
Our event simulation is purely at the parton level which is sufficient for the fully leptonic final state.
We used HELAS \cite{Murayama:1992gi} to calculate the helicity amplitudes for the full process. LHApdf \cite{Whalley:2005nh} was used to fold in the parton distribution functions for the protons using the pdf set CTEQ6L \cite{Pumplin:2002vw}. An adaptive Monte-Carlo package, BASES \cite{Kawabata:1985yt}, was used to perform the integration over phase space and to study differential cross-sections. Our simulation approach to calculation of helicity amplitudes allowed us to insert the form factor specifically into the longitudinal gauge-boson scattering channel.

 The cuts used are: 1): $\Delta r > 0.4$ separation between leptons; 2): $p_T > 20$ GeV and $|\eta| < 2.8$ cuts on the leptons. 3): $Z$-reconstruction cut: We require that two opposite sign leptons reconstruct to give the Z mass. 4): Missing $E_T$ cut $> 20$ GeV. 5): Invariant mass cut of the $W/Z$ system between $m \pm \Gamma$ \footnote{In practice this requires looking for a broad excess and placing an invariant mass cut on the $WZ$. We found that our results were not significantly affected by choosing a different invariant mass cut window, indicating that a precise optimization of this cut may not be very important.}.

When we reconstruct the events, there are two misidentification issues that arise that lead to a fourfold ambiguity in the kinematics: a): The $u$-quark direction is unknown. b): There is a two fold ambiguity in the neutrino momentum along the beam axis. First, consider a misidentification of the $u$-quark direction. This leads to misidentifying $\Theta \rightarrow \pi - \Theta$, $\phi_1 \rightarrow \pi + \phi_1$, $\phi_2 \rightarrow \pi + \phi_2$. The azimuthal angle correlation, $\sin \phi_1$, is odd under such a misidentification. Note that from the solid green curve in Fig.~\ref{fig:coefficients} the coefficient of $\sin \phi_1$ is also approximately odd under $\Theta \rightarrow \pi - \Theta$. Thus, if we study the $\sin \phi_1$ mode for either $0.1<\cos \Theta<0.9$ or $-0.9<\cos \Theta<-0.1$  we find that it is robust to misidentifications of the $u$-quark direction.

The presence of a false solution for the neutrino momentum would distort the azimuthal angle correlations that we seek for $\phi_1$. However, to study the $\sin \phi_1$ mode, we can simply measure the up-down asymmetry with respect to the $\phi_1=0$ (production) plane to find the sizes of such correlations. We will demonstrate that the up-down asymmetry is the same for both the true and the false solutions.

For a given $\phi_1$ azimuthal angle correlation we have,
\begin{equation}
\left . \frac{d \sigma}{d \phi_1} \right | _{ \cos \Theta \gtrless 0} \simeq A_0 + A_1 \cos \phi_1 + A_2 \cos 2\phi_1 \pm B_1 \sin \phi_1 \ .
\end{equation}
We define the events going ``above'' the plane for $\sin \phi_1 > 0$ and going ``below" the plane for $\sin \phi_1 < 0$. Therefore, the up-down asymmetry can be defined as
 \begin{equation}
 \label{eq:AS}
\left . AS \right | _{ \cos \Theta \gtrless 0} = \frac{N_+ -N_-}{N_+ + N_-} = \pm \frac{2}{\pi}\frac{B_1}{A_0} \ ,
\end{equation}
where
$N_+$/$N_-$ are the number of up/down events respectively.

The up or down events for $\phi_1$ can be defined by the sign of the scalar triple product
\begin{equation}
SGN \equiv \textrm{sgn}\left(\hat{n} \cdot p_2 \right) = \textrm{sgn} \left( (k_1 \times q_1). p_2 \right) \ .
\end{equation}
For a particular event if $\textrm{SGN} > 0$$(<0)$ then we increment $N^+$$(N^-)$.
The normal vector to the production plane $\hat{n} = k_1 \times q_1 = k_1 \times (p_1 + p_2)$ is independent of the $\nu$ momentum along the $u$-quark direction and hence $SGN$ is insensitive to the difference between the true and false solution.
In addition to this, the asymmetry variable has the advantage of being insensitive to a number of cuts such as rapidity and $p_T$ cuts that would otherwise distort the angular distribution.

%There is one caveat to the above discussion. In order to identify the sign of $\cos \Theta$, we need to use the neutrino momentum to reconstruct $\cos \Theta$ in the CM frame. For a small class of events, positive $\cos \Theta$ for the true solution will be mapped to $-\cos \Theta$ for the false solution. This would typically result in the true and false solution yielding opposite values of $SGN$. We discard all events for which the true and false solutions yield such opposite values.

%\begin{table}
%\begin{tabular}{|c|c|c|c|}
%  \hline
%  $\Gamma/m \rightarrow$ & 0.2 & 0.3 & 0.4 \\
%  \hline
%  m (GeV)   &  &  &  \\
%  \cline{1-1}
%  1000 & 	5.0 &	5.4 &	5.4 \\
%  1200 &	2.7 &	2.8 &	2.9 \\
%  \hline
%\end{tabular}
%\caption{100 TeV, Table showing the significance of the asymmetry variable, from Eq. (\ref{eq:sig}), for different form-factor parameters in composite Higgs models assuming $a=0.8$ for an integrated luminosity of 3000 fb$^{-1}$. }
%\label{tab:fullresults}
%\end{table}

\smallskip
\noindent{\bfseries Results.} If the background fluctuation is Gaussian, the statistical significance of the nonzero asymmetry is given by,
\bea
\label{eq:sig}
S \equiv \frac  {\left | N^+ - N^-  \right |}{\sqrt{N}} =| AS| \sqrt{N},
\eea
where $N = N^+ + N^-$ is the total number of events. As a rule of thumb, we found that choosing an invariant mass window between $m\pm\Gamma$ seemed to optimize the trade-off between picking up a large $|AS|$ by being close to the resonance, while still keeping a sizeable number of events.

In Tab.~\ref{tab:xsecs}, we show the cross-section for the process under consideration at the 14 TeV LHC and at a future 50~TeV $pp$ collider for different choices of form factors by varying over the parameters $m$ and $\Gamma$.

\begin{table}

\begin{tabular}{|c|c|c|}
\multicolumn{3}{c}{14 TeV LHC}\\
\hline
  $\Gamma/m \rightarrow$  & 0.3 & 0.4 \\
  \hline
  m (GeV)   &  &    \\
  \cline{1-1}
  800 & 		3.4  & 4.8 \\
  900 &         2.4  &  3.3   \\
  1000 & 		1.7 &	2.4 \\
  \hline
\end{tabular}
\begin{tabular}{|c|c|c|c|}
\multicolumn{4}{c}{50 TeV $pp$ collider}\\
  \hline
  $\Gamma/m \rightarrow$ & 0.2 & 0.3 & 0.4 \\
  \hline
  m (GeV)   &  &  &  \\
  \cline{1-1}
  1000 & 	5.7 &	6.7 &	9.1 \\
  1100 &    3.7 &   4.8 &   6.8 \\
  1200 &	2.4 &	3.4 &	5.1 \\
  \hline
\end{tabular}
\caption{Table showing the cross-sections (in fb) for the fully-leptonic process $pp \rightarrow W^\pm Z \rightarrow l^\pm \nu l^+ l^-$  for different form-factor parameters in a composite Higgs model (assuming $a=0.8$) and with the cuts described in the text (including the mass window selection). Cross-sections are shown for the 14 TeV LHC and at a future 50~TeV $pp$ collider. The typical cross-sections are of the order of a few fb.}
\label{tab:xsecs}
\end{table}

The form factors that we consider lead to typical asymmetries of the order of $5-10\%$. In Tab. \ref{tab:fullresults}, we show the significance of the asymmetry measurement at the 14 TeV LHC with 3000 fb$^{-1}$ of data for different choices of form factors by varying over the parameters $m$ and $\Gamma$. We show the results using expected statistics including both the $W^+ Z$ and $W^- Z$ fully-leptonic modes.

We find that new wide resonances can be probed at the $3\sigma$ level for masses up to more than 900 GeV. This further motivates an extended run of the LHC should an excess in $W^\pm Z$ be discovered. A future 50~TeV proton-proton collider could probe resonances up to around $1.2$ TeV with the same luminosity. An even higher energy at a future collider would make the $\sin(\phi_1 + \phi_2)$ interference term the dominant piece and would require a different analysis strategy.

\begin{table}

\begin{tabular}{|c|c|c|}
\multicolumn{3}{c}{14 TeV LHC}\\
\hline
  $\Gamma/m \rightarrow$  & 0.3 & 0.4 \\
  \hline
  m (GeV)   &  &    \\
  \cline{1-1}
  800 & 		4.4 &	4.4 \\
  900 &         3.2 &  3.2 \\
  1000 & 		2.3 &	2.3 \\
  \hline
\end{tabular}
\begin{tabular}{|c|c|c|c|}
\multicolumn{4}{c}{50 TeV $pp$ collider}\\
  \hline
  $\Gamma/m \rightarrow$ & 0.2 & 0.3 & 0.4 \\
  \hline
  m (GeV)   &  &  &  \\
  \cline{1-1}
  1000 & 	4.8 &	5.0 &	4.9 \\
  1100 &    3.3 &   3.6 &   3.6 \\
  1200 &	2.4 &	2.6 &	2.7 \\
  \hline
\end{tabular}
\caption{Table showing the significance of the asymmetry variable, from Eq. (\ref{eq:sig}), using both the $W^\pm Z$ fully-leptonic modes for different form-factor parameters in a composite Higgs model (assuming $a=0.8$). The results are shown for an integrated luminosity of 3000 fb$^{-1}$ at the 14 TeV LHC and at a future 50~TeV $pp$ collider.}
\label{tab:fullresults}
\end{table}

There are several theoretical and analysis issues that could potentially increase these significances in a more sophisticated search and motivates an elaboration of our work. 1): In the $\sin \phi_1$ mode search, it is possible to open up the hadronic decay modes of the $Z$ with boosted tagging techniques \cite{Cui:2010km}. 2): In addition, using a multivariate analysis and incorporating the $\sin \phi_2$ and $\sin(\phi_1+\phi_2)$ modes could also bolster this result. 3): Allowing for $I=2$ resonances implies that the vector resonance form factor could be scaled by a factor larger than $\sqrt{1-a^2}$. 4): Multiple resonances in a narrow mass window could also yield an enhancement in the longitudinal scattering cross-section which would show up as the $P(s)$ factor mentioned earlier.

%\begin{itemize}
%\item  If we focus on the $\sin \phi_1$ mode, then it is possible to open up the hadronic decay modes of the $Z$ with boosted tagging techniques \cite{Cui:2010km}.
%\item In addition using a multivariate analysis and incorporating the $\sin \phi_2$ and $\sin(\phi_1+\phi_2)$ modes could also bolster this result.
%\item Allowing for $I=2$ resonances implies that the vector resonance form factor could be scaled by a factor larger than $\sqrt{1-a^2}$.
%\item Multiple resonances in a narrow mass window could also yield an enhancement in the longitudinal scattering cross-section which would show up as the $P(s)$ factor mentioned earlier.
%\end{itemize}

%\section{Conclusions}

\smallskip
\noindent{\bfseries Conclusions.} We have proposed a novel technique to disentangle the dynamics of a strongly coupled EWSB sector by measuring a phase shift in the the decay azimuthal angle correlations in massive gauge boson scattering. Our results show that a simple up-down asymmetry in leptons from $W$ decay in $pp\rightarrow W^\pm Z$ is robust to a number of event reconstruction ambiguities and is a good probe of broad resonances from strong dynamics. This strongly motivates a high luminosity run of the 14 TeV LHC. A future 50~TeV $pp$ collider could yield conclusive evidence of resonant behavior in the presence of an excess of $WZ$ events at the LHC. Furthermore, we have outlined several analysis strategies and theoretical issues which would significantly increase the reach of searches based upon this technique and could lead to a promising signal at the next run of the LHC.

%\section{Acknowledgments}

\smallskip
\noindent{\bfseries Acknowledgments.} We would like to thank T. Han, D. Krohn, A. Larkoski, and M. Peskin for useful discussions.
H.M. was supported in part by the U.S. DOE under Contract
No. DEAC03-76SF00098, by the NSF under Grant
No. PHY-1002399, by the JSPS Grant (C) No. 23540289,
by the FIRST program Subaru Measurements of Images
and Redshifts (SuMIRe), CSTP, and by WPI, MEXT,
Japan. V.R. was supported by NSF Grant No. PHY-0855561.

 %%%%%%%%%%%%%%%%%%%%%%%%%%%%%

\end{document}